Title:

# Field-driven femtosecond magnetization dynamics induced by ultra-strong coupling to THz transients

**Authors:**
C. Ruchert[1], C. Vicario[1], F. Ardana-Lamas[1,4], P.M. Derlet[2], B. Tudu[3], J. Luning[3] and C.P. Hauri[1,4]

**Affiliations:**
[1]Paul Scherrer Institute, SwissFEL, 5232 Villigen PSI, Switzerland

[2]Paul Scherrer Institute, Condensed Matter Theory Group, 5232 Villigen PSI, Switzerland

[3]Université Pierre et Marie Curie, LCPMR, UMR CNRS 7614, 75005 Paris, France

[4]Ecole Polytechnique Fédérale de Lausanne, 1015 Lausanne, Switzerland

**Abstract:**

**Controlling ultrafast magnetization dynamics by a femtosecond laser is attracting interest both in fundamental science and industry because of the potential to achieve magnetic domain switching at ever advanced speed. Here we report experiments illustrating the ultrastrong and fully coherent light-matter coupling of a high-field single-cycle THz transient to the magnetization vector in a ferromagnetic thin film. We could visualize magnetization dynamics which occur on a timescale of the THz laser cycle and two orders of magnitude faster than the natural precession response of electrons to an external magnetic field, given by the Larmor frequency. We show that for one particular scattering geometry the strong coherent optical coupling can be described within the framework of a renormalized Landau Lifshitz equation. In addition to fundamentally new insights to ultrafast magnetization dynamics the coherent interaction allows for retrieving the complex time-frequency magnetic properties and points out new opportunities in data storage technology towards significantly higher storage speed.**

**Main text:**

Femtosecond coherent manipulation of magnetic moments at ever advanced speed is of paramount interest to future data storage and processing technology in order to meet the growing demand of increased data rates and access speed. Today's widely used mass storage devices, based on magnetic domain flipping by the giant magnetoresistance(GMR) effect (*1,2*), are limited to access times in the nanosecond range. The quest towards faster manipulation of magnetization requires a stimulus which is capable of controlling the magnetization vector more rapidly. Following those lines, femtosecond pulses has been used since 1996 as a precursor to overcome GMR speed limits by inducing faster demagnetization. Laser-induced ultrafast demagnetization has its origin in the heat deposition which causes a paramagnetic phase transition upon crossing the Curie/Neel temperature(*3-5*). The ultrafast heat-induced demagnetization, however, is incoherent and followed by a long-standing thermal recovery phase (up to nanoseconds) which hinders femtosecond re-access. Due to the incoherent nature of the near-infrared stimulus (heat), physical insight and visualization into the ultrafast optomagnetic coupling and the magnetization dynamics occurring at the very onset of the laser-matter interaction has been inaccessible which anticipated a complete understanding of the optomagnetic dynamics. For next generation storage technology, significantly higher impact is expected if heat deposition on the storage media is omitted and magnetization is controlled directly with the stimulus' field. First attempts to circumvent heat-induced demagnetization have been undertaken (*6-9*) but purely field-driven ultrafast dynamics of magnetization remained inaccessible.

In this Letter we report on the control and visualization of femtosecond coherent magnetization dynamics in a ferromagnetic cobalt thin film. An ultrastrong optomagnetic coupling is established by use of an intense single-cycle THz pulse with a stable absolute phase (*10,11*). We give an experimental proof that the THz-induced magnetization dynamics are coherently governed by the absolute phase and amplitude of the initiating magnetic field rather than its intensity envelope and heat. Our studies unveil unexpected rich and ultrafast field-driven magnetization dynamics which are not governed by the (slow) Larmor precession frequency, but by the ultrafast THz stimulus field oscillations. In fact, we show for the first time that the field dynamics of the stimulus are directly imprinted on the magnetization dynamics and that the magnetisation dynamics vanish as soon as the laser field vanishes. The ultrastrong coherent optomagnetic interaction gives for the first time the possibility to retrieve the complex magnetic impulse response in both the time and frequency domain. The experimental data for one particular scattering geometry ($\vec{B}_{THz} \perp \vec{B}_{ext}$) is approximately described by an overdamped Landau-Lifshitz formalism thereby accounting for effects caused by the strong co-propagating electric field. The results represent the dawn of a new era towards purely optical, coherent magnetic domain switching on the time scale of the stimulus' in the complete absence of heat.

As a model system for our investigations, a 10 nm ferromagnetic cobalt thin film at room temperature (290 K) is used, which due to the thin-film shape anisotropy exhibits an in-plane magnetization. Prior to THz excitation, the cobalt macroscopic magnetic moment is well aligned by an external field $\vec{B}_{ext}$ and the magnetization is saturated (Fig. 1). The THz field (blue) with a variable linear polarization is used to initiate optical field-induced coherent magnetization dynamics in the thin film which are detected by the time-resolved magneto-optical Kerr effect (MOKE). As a probe, a sub-50 fs near infrared pulse (λ=800 nm) is employed which is inherently synchronized to the THz pump pulse. The octave-spanning THz pulse carries ≈1.5 optical cycles with a 0.3 Tesla field amplitude, centered at 2.1 THz. Most important for our investigations is the fact that the THz magnetic field $\vec{B}(t)$ is phase-locked to its intensity envelope, $\left|\vec{B}^2(t)\right|$, as is the co-propagating electric field $\vec{E}(t)$. This turns out to be a prerequisite for the observation of sub-cycle magnetization dynamics.

Under approximately equilibrium conditions, the interaction between a magnetic field $\vec{B}$ and the magnetization $\vec{M}$ is well described by the Landau-Lifschitz (LL) equation, $\frac{d\vec{M}}{dt} = -\gamma(\vec{M}\times\vec{B}) - \frac{\gamma\alpha}{|M|}(\vec{M}\times(\vec{M}\times\vec{B}))$. The first term $\vec{M}\times\vec{B}$ leads to simple precession of the magnetic moment $\vec{M}$ around the external field $\vec{B}$, via the Zeeman torque with the gyromagnetic constant γ. This motion is the well-known Larmor precession with the frequency $\nu_L \approx 10\,GHz$ for a 0.3 T magnetic field. The second term leads to a damping of the precession tending to orient the magnetization towards the external field, with the damping factor α an empirical parameter that embodies diverse dissipative processes involving lattice and electronic degrees of freedom. While this picture holds well for a constant or slowly varying external magnetic field, the ultrafast magnetization dynamics triggered by our intense single-cycle THz stimulus are expected to look fundamentally different, in particular owing to the strong co-propagating electric field component ($E_{max}^{THz} \approx 1$ MV/cm).

These expectations are corroborated in figure 2(a). The results show the observed magnetic dynamics represented by the MOKE signal (red line) during the interaction time of the THz magnetic field (black line) with the sample. The THz magnetic field which is orthogonally polarized to the aligned magnetic moment (position "A" in Fig 1) induces a strong and instant torque to the macroscopic magnetic moment. It turns out that the magnetization dynamics are not governed by the Larmor precession frequency caused by the Zeeman interaction term $\gamma \vec{M} \times \vec{B}$. In contrast the magnetic response follows almost instantaneously the much faster oscillations of the strong femtosecond THz pulse featuring frequency components in the multi-THz range. After a delay of ≈50 fs the magnetization moment $\vec{M}$ pursues a de- and re-magnetization cycle similar to the driving THz magnetic field oscillations. The strong coupling of the THz pulse to the magnetization vector is reflected by the large magnetic Kerr rotation of up to 200 mrad measured by MOKE. The MOKE frequency spectrum and its corresponding spectral phase depicted in figure 2b (red line) manifests the occurrence of strong THz constituents with frequencies comparable to the THz stimulus spectrum (black line). The magnetic response is thus fully dominated by the frequency components of the THz magnetic field (2.1 THz and 0.5 THz) while a contribution of the Larmor precession frequency is not observed. Remarkably the strong phase-locked stimulus completely dominates the magnetic dynamics in a reproducible and coherent manner. This fact is reflected in the almost identical spectral phase of both the THz stimulus and the MOKE response (Fig 2b). The presented experimental results clearly manifest that the magnetization $\vec{M}$ follows the initiating magnetic field $\vec{B}(t)$ rather than its intensity envelope $\left|\vec{B}^2(t)\right|$ on a time scale which has not been expected to be that fast (*12*). Our findings clearly corroborate that Terahertz laser pulses with a stable absolute phase offer an entirely new experimental avenue for the exploration and control of coherent sub-cycle magnetization dynamics.

The ultrafast magnetization dynamics become observable thanks to the absence of heating and ionizing effects throughout the interaction. In fact, a Terahertz photon carries a photon energy which is almost three orders of magnitude below that of a near-infrared photon (0.004 eV versus 1.6 eV). This avoids heating of the electronic sub-system and enables the discovery of unexpected rich coherent motions of the magnetization, such as sub-cycle de- and re-magnetization dynamics during the very first optical cycles of the THz pulse interaction. These dynamics are for the first time driven coherently by the phase properties of the intense THz field.

In depth physical insights on the observed dynamics can be gained by further interrogating the LL equation. Even though LL ignores by default the strong electric field contribution of the THz pulse the equation is found to mimic the experimentally observed magnetization dynamics surprisingly well when a large damping term α and modified $H_{ext}$ is introduced. Under those conditions the calculated dynamics of $\vec{M}$ follow the stimulus field oscillations and reproduce a large part of the measured characteristic temporal dynamics (Fig. 2c). As experimentally observed the calculated magnetic moment (green) pursues the stimulus' field oscillations (grey dashed) with a few tens of femtosecond retardation and perfectly reorients in the external field after the THz interaction terminated. A comparison with the experimental data gives evidence that the Zeeman-driven precession is insignificant and dominated by strong damping dynamics whose damping parameter is three orders of magnitude larger than that normally expected for equilibrium Co. While the over-damped LL limit represents well the overall magnetization dynamics (green) with regards to the experimental MOKE signal (red), the subsequent developing phase lag observed in the experiment is less well reproduced indicating some non-linear memory effects as a consequence of the strong electric and magnetic field interaction (13).

The measured magnetization dynamics $\vec{M}(t)$ shown in Fig 2 indicates that the phase properties of the magnetization are fully determined by the phase-locked THz magnetic field. We call this coherent coupling. While conventional heating of the sample above $T_C$ results in a decrease of the macroscopic magnetization magnitude $|\vec{M}|$, the coherent coupling mechanism postulated here is supposed to alter exclusively the angular orientation of $\vec{M}$, but not its net magnitude $|\vec{M}|$. To corroborate this assumption an additional MOKE measurement was performed under identical conditions as shown in Fig. 2a but with an oppositely polarized THz pulse (Fig.1 position "B"). This corresponds to a π shift of the temporal phase. With this inverted coherent stimulus field (Fig. 3(a)) the measured magnetic dynamics unveil a MOKE signal which is equivalently inverted. The sum of the corresponding two MOKE signals leads to a full annihilation of the magnetization dynamics while the difference signal give rise to a two-times enhanced signal (Fig. 3(b)), as expected from a coherent interaction. We emphasize that the annihilated ("zero") signal in (b) is a direct experimental proof for the absence of an incoherent heat stimulus. It is achieved only when the full vectorial (i.e. angular and magnitude) information of M is governed coherently by the THz field leading to a vectorial sum of $\vec{M}$ which is zero. In contrast, an analog experiment with oppositely polarized heat stimuli would yield a sum of the two MOKE signals different to zero since the magnitude $|\vec{M}|$ of the magnetization undergoes a change via the paramagnetic phase transition caused by heating. In addition the high electronic temperature smears out the orientation of the previously well aligned magnetic moment and introduces incoherent motion to the magnetic dynamics. Those two contributions would result in a vector sum of the corresponding MOKE signals unequal to zero. A direct control on the angular orientation of $\vec{M}$ without changing its magnitude is thus impossible for a heat-based stimulus, and hinders coherent control on magnetization dynamics. We therefore conclude that our results represent for the first time unambiguously a coherent and ultrafast coupling of the strong THz magnetic field to the magnetization moment with the complete absence of heat.

It is this coherent nature of interaction which gives for the first time access to the complex magnetic response (i.e. complete information on phase and amplitude), both in the time and frequency domain. As is known from signal processing (Fig. 4a) the impulse response, $\Re(t)$, of a system is determined by the measured input, I(t), and output, $\Omega(t)$, and can be reconstructed in the time domain by means of deconvolution of I and $\Omega$ or likewise by inverse Fourier transformation of the spectral response function $\widetilde{\Re}(\omega)$. In our experiment the complex response function $\widetilde{\Re}(\omega) = \widetilde{I}(\omega)/\widetilde{\Omega}(\omega)$ has been reconstructed following the procedure laid out in Fig 4a by transforming the measured complex temporal THz field I(t) and MOKE signal $\Omega(t)$ to the frequency domain. The spectral amplitude response $\Re(\omega)$ and the corresponding spectral phase $\phi_\Re$ calculated from the experimental data shown in Fig 2(b), are presented in Fig 4b and c (red curve). The spectral response, far from being flat, unveils the dominant frequencies of the cobalt thin film in response to illumination by our broadband THz stimulus. The magnetic system unveils a strong response in the lower frequency part (<1 THz) which fades out towards higher frequencies. For sake of completeness the corresponding temporal impulse response of our ferromagnetic sample is shown as an inset in Figure 4(b).

Of particular interest is the narrow-band frequency response at 1 THz disclosed by our time and frequency resolved measurements. Its origin is allocated to the strong co-propagating terahertz electric field (1 MV/cm) which is supposed to influence the magnetic response measurably. In order to corroborate this statement we performed an additional MOKE measurement. This time the magnetic field polarization was set parallel to the external field ($\vec{B} \parallel \vec{B}_{ext}$) in order to disable magnetic coupling according to LL (position "C", Fig. 1). The MOKE frequency spectrum recorded for this configuration (Fig. 4(c), black) is surprisingly still present yet significantly different to what is observed for the Zeeman active (i.e. $\vec{B}_{THz} \perp \vec{B}_{ext}$) configuration "A" (4c, red dotted line). In the latter the frequency components around 1 THz are barely present, nor are higher frequencies (> 2.5 THz), which hint at a different coupling mechanism for the two configurations. It is, however, remarkable that for both field configurations the frequency impulse response carries the narrow-band signature at 1 THz with an associated jump in the spectral phase (Fig 4b).

We propose that the observed magnetic response could be associated to the current induced by the strong electric field according to Ampere circuital law. The corresponding response functions give evidence that this coupling is present in both configuration $\vec{B} \parallel \vec{B}_{ext}$ and $\vec{B} \perp \vec{B}_{ext}$, though dominated by the THz magnetic field induced dynamics for the latter. While the origin of this phenomenon requires further studies our investigations unveil that the strong co-propagating electric field has an observable effect on the magnetization dynamics. Since the standard LL equation does not consider strong co-propagating electric fields the above mentioned phenomenon cannot be described adequately. A more general theory is needed to give a complete picture of the magneto-electric dynamics in this strong-field electro-magnetic regime which includes a nonlinear integro-differential approach for LLG coupled to the Maxwell equations, as developed for magneto-electric materials (*13*). Our numerical investigations (Fig. 2c) show, however, that for the case of $\vec{B}_{THz} \perp \vec{B}_{ext}$, the complex nonlinear formalism proposed in ref. 14 can be transposed into an effective linear LL by a simple renormalization which can describe reasonably well the magnetization dynamics observed in our thin film.

In former studies there has been evidence that heat-induced demagnetization with near-infrared lasers (hν≈1.6 eV) alters the dielectric tensor significantly and thus the MOKE signal due to excitation of electrons (*4*). In those experiments the illumination of magnetic thin films give rise to hot electrons (typically ≥1000 K) and thus dichroic bleaching (*14,15)* which obviates a detailed response of the MOKE signal within the first tens of femtoseconds. The use of a THz pump laser prevents such deceptive MOKE signals since the creation of hot electrons by THz is virtually excluded and so are state-blocking transitions. Our calculation shows that the THz interaction with the electronic system gives rise to a temperature increase of the electrons as low as a few tens of Kelvin above room temperature thanks to the low THz photon energy (see *supplementary information)*. The reported MOKE signal reflects therefore the unbiased and coherent magnetic dynamics in the cobalt thin film. Finally we would like to mention that the total amount of angular moment carried by the THz photons is approximately three orders of magnitudes larger than for a conventional 800 nm stimulus with equal pulse energies. This, and the absence of heat during the interaction makes us believe that intense THz pulses are an important new tool to investigate the magnetization dynamics and to shed light onto the not yet fully understood interaction mechanism of angular momentum between light and matter during magnetic switching.

In conclusion, the new aspects of our investigations involve the observation of coherent sub-cycle femtosecond magnetization dynamics in ferromagnetic thin films controlled by a strong single-cycle THz stimulus. In the absence of heat injection the magnetization vector is shown to be controlled coherently by the complex field of the THz laser, determined by its amplitude and phase. Previously inaccessible sub-cycle magnetization dynamics are therefore visualized and are shown to be governed by the frequency components of the THz pump rather than the natural response frequency of the electrons (Larmor frequency). The coherent, phase-sensitive coupling of the THz field to the magnetization allows for the retrieval of the complex impulse response of the magnetic thin film, both in the time and frequency domain. The presented concept of a phase-stabilized non-ionizing stimulus opens the door towards coherent magnetic domain switching in absence of heat towards the yet unknown speed limit of magnetization dynamics.

**Acknowledgments:**

This work was carried out at the Paul Scherrer Institute and was supported by the SNF grant PP00P2_128493 and SwissFEL. We acknowledge M. Paraliev for valuable support. B.T. is supported by the ERASMUS Mundus program. J.L. acknowledges the DYNAVO project for financial support for upgrading the magnetron sputtering.

**Correspondence:**

Correspondence and requests for materials should be addressed to C.P.H.

(christoph.hauri@psi.ch;christoph.hauri@epfl.ch)

**Author contributions:**

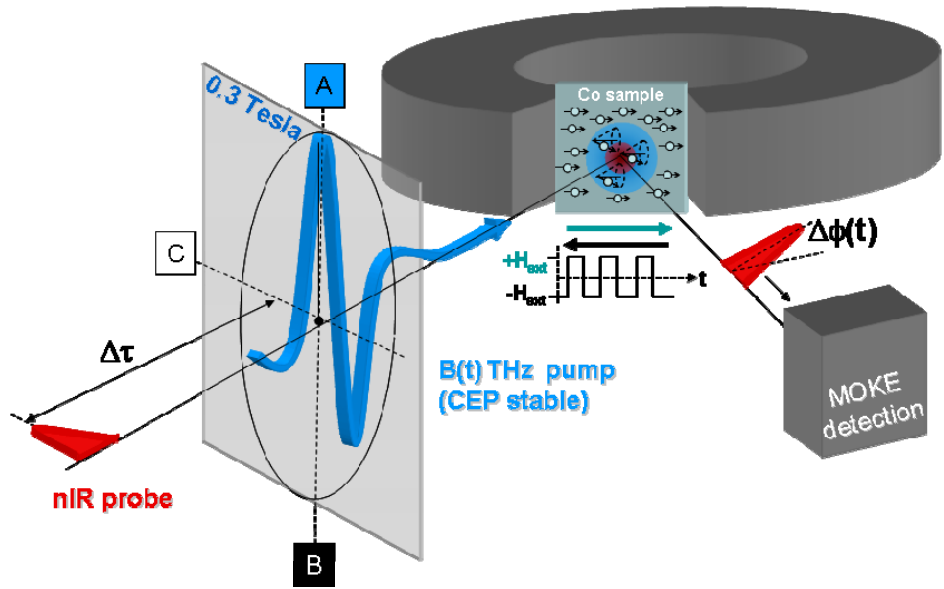

**Fig. 1**. Experimental scheme with time-resolved magneto-optical Kerr effect (MOKE) that allows for the measurement of ultrafast moment dynamics on the Co sample surface kept at room temperature. The strong 0.3 T single-cycle THz magnetic field (blue) is linearly polarized and carries an absolute phase which is constant for consecutive shots. The polarization can be rotated by 90 degree (position C) and inverted (position B), corresponding to a π phase shift. The co-propagating electric field is not shown. The THz pump pulse hits the sample under 20 degree off normal incidence. The collinear 50-fs probe pulse originates from the same laser system driving the THz source and allows for jitter free time-resolved measurement of magnetization dynamics by MOKE. The sample is placed in an external DC magnetic field ($\vec{B}_{ext} = 0.1\,\text{T}$) which can be inverted after a pump-probe cycle (by inverting the magnet current) in order to improve the signal-to-noise of the Kerr rotation.

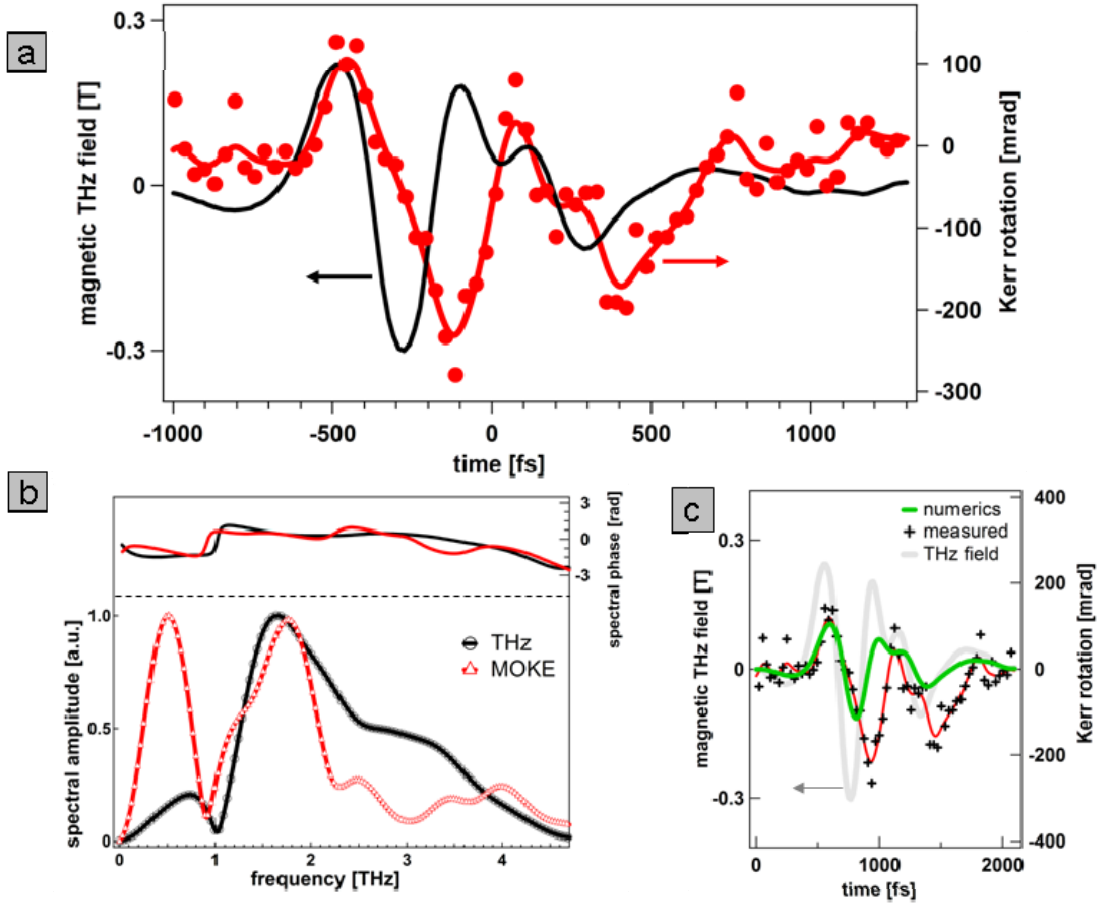

**Fig. 2**. Magnetization dynamics initiated by the strong THz transient. (a) Femtosecond magnetization dynamics represented by the MOKE signal (red) and initiated by the phase-locked single-cycle THz magnetic field (black). The magnetization follows almost instantaneously the THz field oscillations and its dynamics is governed by frequencies much higher than the Larmor contribution (0.01 THz). The corresponding spectra and spectral phases are shown in (b) for the THz pulse (black) and the magnetization dynamics (red). Interrogating Landau-Lifshitz (LL) equation provides insight into ultrafast dynamics shown in (c). A renormalized, strongly damped LL (green) mimics the overall temporal magnetization dynamics surprisingly well when compared to the experimental MOKE results (red, from (a)) even though LL ignores by default the co-propagating electric field.

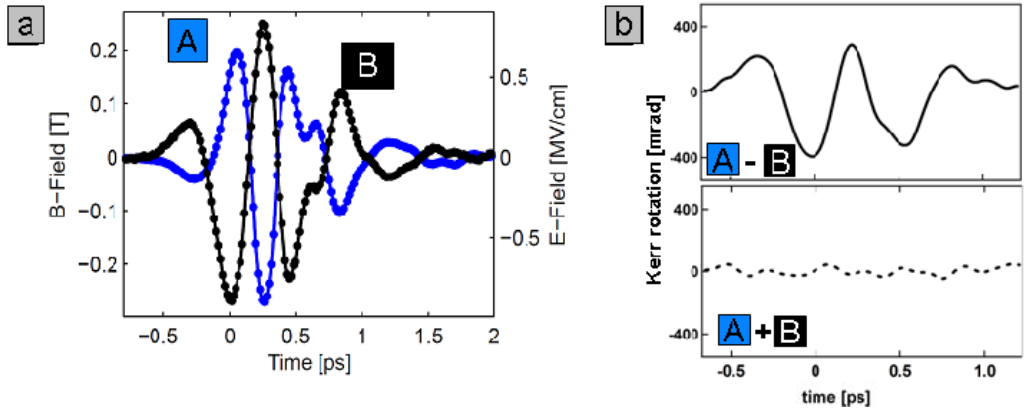

**Fig. 3**. Investigation of coherence properties of the initiated magnetization dynamics. (a) The fundamental THz stimulus linearly polarized in direction "A" and the identical, but inverted ($\pi$-phase shifted) copy "B" used to initiate magnetization dynamics under otherwise equal conditions. (b) The calculated difference and sum signal of the measured MOKE traces show a two times amplification and complete annihilation, respectively. These findings are an unambiguous proof for the absence of an incoherent heat-based stimulus. The absence of an incoherent heat stimulus allows for the first time a fully coherent steering of the magnetization vector thanks to the coherent coupling of the THz field to the magnetization moment.

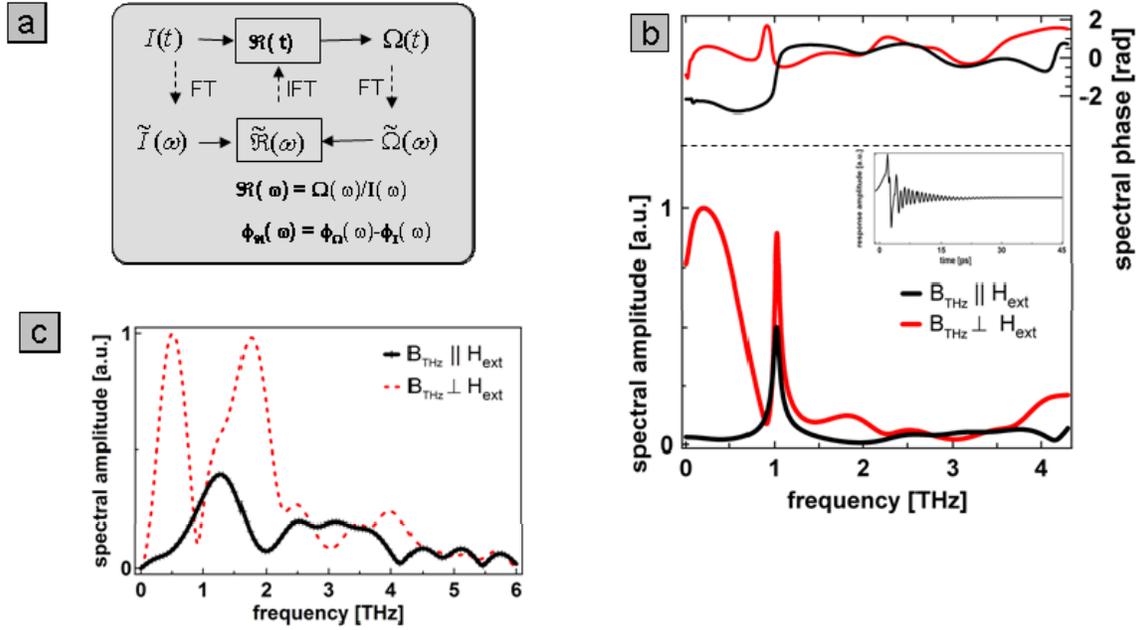

**Fig. 4**. The complex impulse response of the cobalt thin film. (a) Fourier transformation of the measured THz field I(t) and MOKE signal $\Omega(t)$ gives access to the complex functions $\tilde{I}(\omega)$ and $\tilde{\Omega}(\omega)$ which allow direct calculation of the complex response function $\tilde{\Re}(\omega)$ shown in (b) with spectral amplitude $\Re(\omega)$ and spectral phase $\Phi_{\Re}(\omega)$ (both in red). The temporal response function $\Re(t)$ is gained by inverse Fourier transformation and shown in the inset of (b). The frequency response function exhibits a dominant broadband frequency response centered around 0.4 THz and a resonance frequency at 1.05 THz, which is attributed to the strong co-propagating THz electric field. This assumption is corroborated by the observed MOKE signal measurement with the THz B field parallel to the initial magnetization (c, black) which is significantly different to the Zeeman configuration (red dotted). The complex magnetic response function for this configuration (b, black line) consists only of the narrowband resonant mode at 1.05 THz. Since no direct coupling of the magnetic THz field component to the magnetization is present the observed magnetic response is attributed to the strong pulsed co-propagating electric field.

**Supplementary Materials:**

Methods:

Laser-based THz source and pump-probe setup

The single-cycle THz pulses are generated by optical rectification in the organic crystal DAST (4-N,N-dimethylamino-4'-N'-methyl stilbazolium tosylate). The output of a Ti:sapphire amplifier system delivering up to 20 mJ, 50 fs pulses at 100 Hz is used to pump a whitelight-seeded optical parametric amplifier (OPA) system with three amplifier stages. The OPA delivers 45fs, 2.5 mJ pulses at a wavelength of 1.5 µm which are used to initiate optical rectification in the nonlinear organic crystal. Frequency conversion with an efficiency of ≈2% is achieved in a collimated pump geometry and the collinearly generated THz is redirected to the cobalt sample after separated of the residual pump light. The broadband, single-cycle THz pulses are isolated from the infrared pump laser by a few mm thick teflon filter. No residual infrared (1.5 µm) is detected at the sample position. The THz source offers intrinsically carrier envelope phase-stable (CEP) pulses with a high energy stability of <1%. The temporal and spectral characteristics of the THz field are reconstructed by electro-optical sampling in a 100 µm thick Gallium Phosphide (GaP) crystal. The broadband THz spectrum supports the formation of close to transform-limited single-cycle pulses at a central frequency of 2 THz. The radiation is linearly polarized and the direction of the polarization can be continuously varied by rotating the input polarization and correspondingly the OR crystal. Gold coated mirrors are used throughout the setup for THz transport. Time-resolved magneto-optical Kerr effect (MOKE) experiments are performed by splitting the probe beam from the fundamental laser (Ti:Sapphire) prior to the optical parametric amplifier which is used for pumping the THz source. Both beams (THz and near IR) are thus intrinsically synchronized and hit the sample collinear at close to normal incidence (≈20 deg from the normal). The near IR probe spot size (150 μm at $1/e^2$) is chosen to be significantly smaller than the THz spot size (800 μm at $1/e^2$) to ensure homogeneous Terahertz fields for the MOKE measurements. The absorbed THz pump fluence on the sample is 0.8 mJ/cm$^2$ and thus considerably lower compared to recent near IR pump experiments applying 10-100 mJ/cm$^2$ (*9*). The MOKE system has been identified to be free from artifacts which could potentially arise from the infrared (1.5 μm) pump. The MOKE signal disappeared completely when the THz production was terminated by detuning the DAST from optical rectification. The Kerr rotation is calculated from the polarization rotation of the optical probe ($\lambda_{probe}$=800 nm).

Sample preparation

The 10 nm thick cobalt sample is covered with a 3 nm Al layer and evaporated onto a Si/Pd base substrate covered with 2 nm of Palladium. The aluminum layer on the top is virtually transparent to the THz and the probe pulse. Its hcp crystalline structure exhibits an in-plane isotropy with the c-axis normal to the surface.

Electron temperature during interaction

The total energy of the electron, spin and the lattice will increase during interaction with the laser. Following the model in ref (*16*) we assume the existence of three thermalized reservoirs with temperature $T_e$, $T_s$, $T_l$ that exchange energy. The evolution of the system in time is given by the following three differential equations:

$$C_e(T_e)\frac{dT_e}{dt} = -G_{el}(T_e - T_l) - G_{es}(T_e - T_s) + P(t)$$

$$C_s(T_s)\frac{dT_s}{dt} = -G_{es}(T_s - T_e) - G_{sl}(T_s - T_l)$$

$$C_l(T_l)\frac{dT_l}{dt} = -G_{el}(T_l - T_e) - G_{sl}(T_l - T_s)$$

with $C_e$, $C_s$, $C_l$ the electronic, magnetic, lattice contribution to the specific heat, respectively and $G_{el}$, $G_{es}$, $G_{sl}$ the electron-lattice, electron-spin, and spin-lattice interaction constants and P(t) the instantaneous laser power. For comparison the evolution of the three corresponding temperatures are plotted for two different wavelengths, namely $\lambda_{nIR}$=0.8 µm and $\lambda_{THz}$=150 µm assuming equal fluence on the sample (0.8 mJ/cm$^2$). The results plotted in Fig 5 indicate that the electronic temperature in the case of illumination with nIR is significantly higher than for a THz pump pulse. For THz radiation the temperature increase in the electronic system is approximately 50 K and thus very small. The equivalent electronic temperature for a near IR pump pulse exceeds 1100 K which leads to the state-blocking and dichroic bleaching effects described in ref (*4*).

For sake of completeness we add the calculated penetration depth L given by $L = \lambda / 4\pi k$ with k the imaginary part of the complex refractive index $n = n' + ik$. For the probe beam (k=4.8, λ=800nm) a penetration depth of 13 nm is calculated while for the

THz pump L is 28 nm (k=416, λ=150 μm), both calculated for normal incidence geometry. The k values are taken from (*17*) and extrapolated to the THz range.

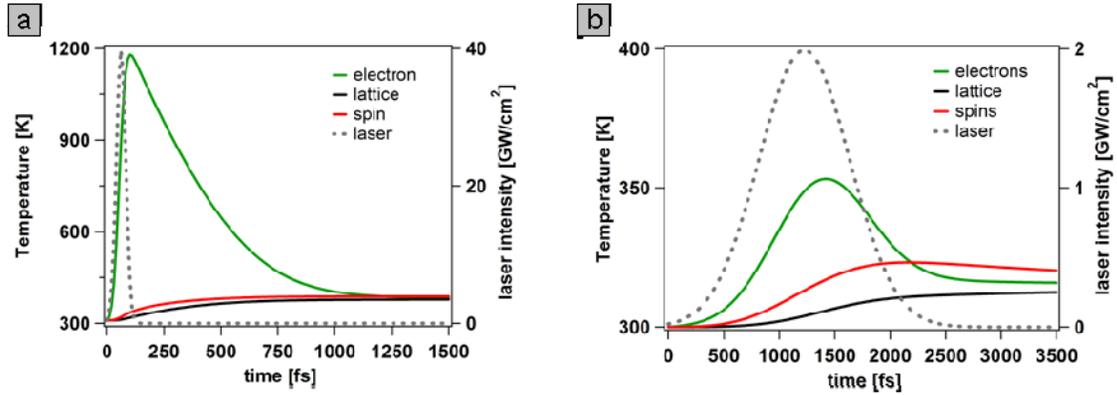

**Fig. 5**. Evolution of electronic, lattice and spin temperatures after illumination of a nearIR pulse (a) and a THz pulse (b) with same fluence (0.8 mJ/cm$^2$). The maximum electron temperature increase reached by THz induction is ≈350 K, thus only slightly above room temperature (RT). For nIR illumination the maximum electron temperature is ≈1200 K, a factor of ≈4 above RT. Different to experiments with near infrared stimuli (λ=800 nm), the THz pump pulse does not significantly alter the electron temperature and provides thus an unbiased insight into magnetization dynamics by means of the time-resolved magneto-optical Kerr effect.

Landau-Lifschitz-Gilbert Equation

The Landau-Lifshitz equation (*18*) is given by

$$\frac{d\vec{M}}{dt} = -\gamma \left(\vec{M} \times \vec{B}_{eff}\right) - \frac{\gamma\alpha}{M_0}\left(\vec{M} \times \left(\vec{M} \times \vec{B}_{eff}\right)\right),$$

where $\vec{M}$ is the magnetization vector with magnitude $M_0$, $\vec{B}_{eff}$ is an external magnetic field and $\gamma$ the gyromagnetic ratio (T$^{-1}$s$^{-1}$). For the present simulations $\vec{B}_{eff}$ consists of the sum of the rapidly varying THz pulse and the static external field ($\vec{B}_{ext} = 0.1\,\text{T}$). In the above equation, $\gamma|\vec{B}_{eff}|$ represents the instantaneous Larmor precession frequency and $\alpha$ is a dimensionless empirical damping parameter that sets the relaxation time-scale via $\left(\alpha\gamma|\vec{B}_{eff}|\right)^{-1}$. For Co under equilibrium conditions, $\alpha \sim 0.014$. Use of these parameter values would lead to a magnetization dynamics dominated by precession around, and relaxation to, the static external field at a time-scale that is in the nano-second range. To produce the femtosecond time evolution entailed in fig. 2c, and approximate agreement with experiment for the case of the THz magnetic component being perpendicular to the static external field, $\alpha \sim 50$ and $\vec{B}_{ext} = 2\,\text{T}$, the latter of which is aligned along the MOKE probe direction. In this over-damped regime the magnetization then approximately follows the THz signal.